\documentclass[12pt]{article}
\usepackage{amssymb}
\usepackage{epsfig}
\usepackage{float}
\usepackage{graphicx}

\begin{document}

\begin{center}
 {\bf Neutrino oscillations and uncertainty relations }
\end{center}

\begin{center}
S. M. Bilenky
\end{center}

\begin{center}
{\em  Joint Institute for Nuclear Research, Dubna, R-141980,
Russia\\}

and

{\em Physik-Department E15, Technische Universit\"at M\"unchen,\\
D-85748 Garching, Germany}

\end{center}
\begin{center}
F. von  Feilitzsch and W. Potzel
\end{center}
\begin{center}
{\em Physik-Department E15, Technische Universit\"at M\"unchen,\\
D-85748 Garching, Germany}
\end{center}

\begin{abstract}
We show that coherent flavor neutrino states are produced (and detected) due to the momentum-coordinate Heisenberg uncertainty relation. The Mandelstam-Tamm time-energy uncertainty relation requires non-stationary neutrino states for oscillations to happen and determines the time interval (propagation length) which is necessary for that. We compare different approaches to neutrino oscillations which are based on different physical assumptions but lead to the same expression for the neutrino transition probability in standard neutrino oscillation experiments. We show that a M\"ossbauer neutrino experiment could allow to distinguish different approaches and we present arguments in favor of the $^{163}$Ho - $^{163}$Dy system for such an experiment.
\end{abstract}

\section{Introduction}

The observation of neutrino oscillations in atmospheric \cite{SK}, solar \cite{Solar}, reactor \cite{Kamland} and accelerator experiments \cite{K2K,Minos} is
one of the most important recent discoveries in particle physics. Small neutrino masses can not be of Standard-Model origin and are commonly considered as a signature of new physics  beyond the Standard Model.

All existing neutrino-oscillation data with the exception of the data of the LSND \cite{LSND} and MiniBooNE antineutrino experiments \cite{Miniboone}, which require confirmation, are perfectly described under the assumption of three-neutrino mixing
\begin{equation}\label{3numix}
\nu_{lL}(x)=\sum^{3}_{i=1}U_{li}\nu_{iL}(x).
\end{equation}
Here $U$ is the PMNS \cite{BPont1,MNS} $3\times 3$ mixing matrix, which is characterized by three mixing angles $\theta_{12}, \theta_{23}, \theta_{13}$ and the $CP$ phase $\delta$, $\nu_{i}(x)$ is the field of neutrinos (Dirac or Majorana) with mass $m_{i}$, and the "mixed field" $\nu_{lL}(x)$ is the SM field which enters into the standard charged current
\begin{equation}\label{standCC}
j _{\alpha}(x)=\sum_{l=e,\mu,\tau}\bar \nu_{l L}(x) \,\gamma_{\alpha}\, l_{L}(x).
\end{equation}
Existing neutrino-oscillation data are analyzed under the assumption that the transition probabilities between different flavor neutrinos are given by the following standard expression (see, for example, \cite{BGG})
\begin{equation}\label{probabil}
\mathrm{P}(\nu_{l}\to \nu_{l'})=\delta_{l'l}-2~\mathrm{Re}\sum_{i>k}U_{l'i}U^{*}_{li}U^{*}_{l'k}U_{lk}
(1-e^{-i\frac{\Delta m^{2}_{ki}L}{2E}}).
\end{equation}
Here, $L$ is the distance between neutrino source and neutrino detector, $E$ is the neutrino energy, $\Delta m^{2}_{ki}= m^{2}_{i}-m^{2}_{k}$.
Notice that it is also convenient to use for the transition probability another expression\footnote { In order to derive this expression we extracted from the  transition amplitude
the common phase factor $e^{-i\frac{m^{2}_{j}L}{2E}}$, where the index $j$ can take the value 1, 2 or  3.
Thus, it is evident that the {\em transition probability }
does not depend on the fixed index $j$.}
\begin{eqnarray}\label{probabil1}
&& \mathrm{P}(\nu_{l}\to \nu_{l'}) = \delta_{l'l}-2\sum_{i}|U_{li}|^{2}(\delta_{l'l}-|U_{l'i}|^{2})(1-\cos\frac{\Delta m^{2}_{ji}L}{2E}) \\\nonumber
  && +2~\mathrm{Re}\sum_{i>k}U_{l'i}U^{*}_{li}U^{*}_{l'k}U_{lk}
(e^{-i\frac{\Delta m^{2}_{ji}L}{2E}}-1)(e^{i\frac{\Delta m^{2}_{jk}L}{2E}}-1),
\end{eqnarray}
where the index $j$ is fixed.

The character of neutrino oscillations is determined by the following two observed features of the neutrino-oscillation parameters:
\begin{itemize}
  \item The solar-KamLAND mass-squared difference
  $\Delta m^{2}_{S}$
is much smaller  than the atmospheric-accelerator mass-squared difference $\Delta m^{2}_{A}$:
 \begin{equation}\label{probabil2}
\Delta m^{2}_{S}\simeq \frac{1}{30}~\Delta m^{2}_{A}.
 \end{equation}
  \item
The mixing angle $\theta_{13}$ is small \cite{Chooz}:
\begin{equation}\label{probabil3}
 \sin^{2} \theta_{13}\leq 4\cdot 10^{-2}.
\end{equation}
\end{itemize}

From (\ref{probabil1}), (\ref{probabil2}) and (\ref{probabil3}) follows (see, for example, \cite{BGG}) that the leading oscillations in the atmospheric and accelerator experiments  are $\nu_{\mu}\rightleftarrows \nu_{\tau}$ and $\bar\nu_{\mu}\rightleftarrows \bar\nu_{\tau}$ and in solar and KamLAND experiments the leading oscillations are $\nu_{e}\rightleftarrows \nu_{\mu,\tau}$ and $\bar\nu_{e}\rightleftarrows \bar\nu_{\mu,\tau}$.

In the leading approximation it is impossible to distinguish two possible neutrino mass spectra:
\begin{itemize}
  \item Normal spectrum
   $$m_{1}< m_{2}< m_{3},\quad \Delta m^{2}_{12}\ll \Delta m^{2}_{23}.$$
   \item Inverted spectrum
 $$m_{3}< m_{1}< m_{2},\quad \Delta m^{2}_{12}\ll |\Delta m^{2}_{13}|.$$
\end{itemize}
In the case of the normal spectrum $\Delta m^{2}_{12}=\Delta m^{2}_{S},~~\Delta m^{2}_{23}=\Delta m^{2}_{A}$ and in the case of the inverted spectrum
$\Delta m^{2}_{12}=\Delta m^{2}_{S},~~\Delta m^{2}_{13}=-\Delta m^{2}_{A}$.

From the recent three-neutrino analysis of the Super-Kamiokande data \cite{SK}
the following 90\% CL limits were found for  the normal (inverted) neutrino mass spectrum
\begin{equation}\label{SK1}
1.9~ (1.7)\cdot10^{-3}\leq \Delta m^{2}_{A}\leq 2.6~(2.7)\cdot 10^{-3}~\rm{eV}^{2},\quad    0.407\leq \sin^{2} \theta_{23}\leq 0.583.
\end{equation}
For the parameter $\sin^{2} \theta_{13}$ the following bounds were obtained
\begin{equation}\label{SK2}
    \sin^{2} \theta_{13}\leq 4\cdot 10^{-2}~ (9\cdot 10^{-2}).
\end{equation}
From the two-neutrino analysis of the MINOS data
was found \cite{Minos}
\begin{equation}\label{Minos}
\Delta m_{A}^{2}=(2.43 \pm 0.13)\cdot 10^{-3}~\rm{eV}^{2},\quad
\sin^{2}2 \theta_{23}>0.90
\end{equation}
From the three-neutrino global analysis of the solar and  reactor KamLAND data
was obtained \cite{Kamland}
\begin{equation}\label{Kamland1}
\Delta m_{S}^{2}=(7.50^{+0.19}_{-0.20})\cdot 10^{-5}~\rm{eV}^{2},\quad
\tan^{2}\theta_{12}=0.452^{+0.035}_{-0.032}
\end{equation}
For the parameter $\sin^{2}\theta_{13}$ was found
\begin{equation}\label{Kamland2}
\sin^{2}\theta_{13}=0.020^{+0.016}_{-0.018}
\end{equation}
At present four neutrino-oscillation parameters ($\Delta m_{S}^{2}$, $\Delta m_{A}^{2}$, $\sin^{2}2 \theta_{23}$ and $\tan^{2}\theta_{12}$) are known with accuracies within the (3-10)\% range. In the accelerator neutrino oscillation experiment T2K \cite{T2K} the parameter $\Delta m_{A}^{2}$ will be measured with an accuracy of $\delta\Delta m_{A}^{2}<10^{-4}~\mathrm{eV^{2}}$ and the parameter $\sin^{2}2\theta_{23}$ will be measured with an accuracy of $\delta(\sin^{2}2\theta_{23})\simeq 10^{-2}$. One of the major aims of this experiment and the reactor experiments
DOUBLE CHOOZ\cite{Dchooz}, RENO \cite{Reno}, and Daya Bay \cite{Dayabay} is to determine the value (or to improve the upper bound by one order of magnitude or better) of the parameter $\sin^{2}\theta_{13}$. In case that this parameter is relatively large, it is envisaged that in future neutrino experiments the value of the $CP$ phase $\delta$ will be determined and the problem of the neutrino mass spectrum will be resolved (see \cite{T2K}).

Thus,  we are entering into the  era of high precision neutrino oscillation experiments.
Despite that the neutrino oscillation formalism, on which the analysis of experimental data is based, has been developed and debated in many papers starting from the 1970s (see reviews \cite{BPont,BPetcov}), these debates and discussions are continuing (see recent papers \cite{Oscillations}). From our point of view the importance of uncertainty relations was not sufficiently analyzed in previous discussions.  We will show  here that the phenomenon of neutrino oscillations is heavily based on the Heisenberg uncertainty relation and the Mandelstam-Tamm time-energy uncertainty relation. We briefly consider different approaches to neutrino oscillations and discuss a M\"ossbauer neutrino experiment which could allow to distinguish them.

\section{Flavor neutrinos: production, evolution,\\
detection}

Which neutrino states are produced in CC weak processes together with charged leptons in the case of neutrino mixing, eq.(\ref{3numix}): Neutrino flavor states, coherent superpositions of plane waves, or
superpositions of wave packets? Here we will present arguments based on the QFT, the Heisenberg uncertainty relation and the knowledge of the neutrino mass-squared differences
that "mixed" flavor states which describe the flavor neutrinos $\nu_{e}$, $\nu_{\mu}$ and $\nu_{\tau}$ are physical states (fully analogous to the "mixed" states which describe  $K^{0}$ and $\bar K^{0}$, $B^{0}$ and $\bar B^{0}$, etc.).

Let us consider (in the lab. system) the decay \cite{BilGiun}
\begin{equation}\label{decay}
 a\to b +l^{+}+\nu_{i} ,\quad (i=1,2,3)
\end{equation}
where $a$ and $b$ are some hadrons.

The state of the final particles is given by
\begin{equation}\label{decay1}
|f\rangle=\sum_{i}|\nu_{i}\rangle|l^{+}\rangle|b~\rangle
\langle \nu_{i} ~l^{+}b|S|a\rangle,
\end{equation}
where $\langle \nu_{i} ~l^{+}b|S|a\rangle$ is the matrix element of the transition
$a\to b +l^{+}+\nu_{i}$, $|\nu_{i}\rangle$ is the state of a neutrino with mass $m_{i}$, momentum $\vec{p}_{i}=p_{i}\vec{k}$ ($\vec{k}$ is the unit vector) and helicity equal to -1. We assume, as usual, that initial and final particles have definite momenta.

The neutrinos $\nu_{i}$ differ only by their masses. If $m_{i}=m_{k}$, in this case $p_{i}=p_{k}$. Taking  into account this requirement for ultra relativistic neutrinos, from general dimensional arguments we have
\begin{equation}\label{decay2}
|p_{i}-p_{k}|\simeq \xi\frac{|\Delta m_{ik}^{2}|}{2E}=\xi\frac{2\pi}{L^{ik}_\mathrm{osc}},
\end{equation}
where $E$ is the energy of neutrinos for $m_{i}^{2}\to 0$, $\xi$ is a coefficient  of the order of one (which can depend on the production process; see \cite{CGiunti}) and
\begin{equation}\label{osclength}
L^{ik}_\mathrm{osc} =4\pi~\frac{E}{\Delta m_{ik}^{2}}\simeq 2.48~\frac{(E/\mathrm{MeV)}}{(\Delta m_{ik}^{2}c^{4}/\mathrm{eV}^{2})}~\mathrm{m} 
\end{equation}
is the oscillation length. For $E\simeq 1$ GeV and $\Delta m_{A}^{2}\simeq 2.4\cdot 10^{-3}~\mathrm{eV}^{2}$ (atmospheric and LBL accelerator neutrinos) we have $L^{A}_\mathrm{osc}\simeq 10^{3} $ km. For $E\simeq 3$ MeV and $\Delta m_{S}^{2}\simeq 7.5\cdot 10^{-5}~\mathrm{eV}^{2}$ (reactor antineutrinos) we have
$L^{S}_\mathrm{osc}\simeq 10^{2} $ km.

On the other side, from the Heisenberg uncertainty relation we have
\begin{equation}\label{decay4}
(\Delta p)_{\mathrm{QM}} \simeq \frac{1}{d}.
\end{equation}
Here $d$ characterizes the quantum-mechanical size of the source. Taking into account that
\begin{equation}\label{decay5}
 L^{A,S}_\mathrm{osc}\gg d
\end{equation}
we have
\begin{equation}\label{decay6}
|p_{i}-p_{k}|\ll (\Delta p)_{\mathrm{QM}}.
\end{equation}
Thus, we conclude that due to the uncertainty relation it is impossible to resolve the emission of neutrinos with different masses.

The operator
\begin{equation}\label{matelem1}
\sum_{k} U^{*}_{lk}\bar\nu_{kL}(x) \gamma_{\alpha}l_{L}(x)
\end{equation}
determines the leptonic part of the matrix element of the process (\ref{decay}). We have
\begin{equation}\label{matelem2}
U^{*}_{li}~\bar u_{L}(p_{i})\gamma_{\alpha}u_{L}(-p_{l})\simeq
U^{*}_{li}\bar u_{L}(p)\gamma_{\alpha}u_{L}(-p_{l}),
\end{equation}
where $p_{l}$ is the momentum of $l^{+}$, and $p=E$ is the momentum of the neutrino for $m_{i}^{2}\to 0$. For the total matrix element of the process
(\ref{decay}) we have
\begin{equation}\label{matelem3}
\langle \nu_{i}~l^{+}b|S|a\rangle\simeq U^{*}_{li}~\langle \nu_{l}~l^{+}b|S|a\rangle_{SM},
\end{equation}
where $\langle \nu_{l}~l^{+}b|S|a\rangle_{SM}$ is the Standard Model matrix element of the emission of the flavor neutrino $\nu_{l}$
with the momentum $p$ in the process
\begin{equation}\label{matelem4}
 a\to b +l^{+}+\nu_{l}.
\end{equation}
From (\ref{decay1}) and (\ref{matelem3}) we find
\begin{equation}\label{matelem5}
|f\rangle=|\nu_{l}\rangle|l^{+}\rangle |b~\rangle
\langle \nu_{l}~l^{+}b|S|a\rangle_{SM},
\end{equation}
where {\em the state of the flavor neutrino $\nu_{l}$} is given by the relation
\begin{equation}\label{matelem6}
|\nu_{l}\rangle=\sum_{i}U^{*}_{li}~|\nu_{i}\rangle \quad (l=e,\mu,\tau)
\end{equation}
and $|\nu_{i}\rangle $ is the state of a neutrino with mass $m_{i}$, negative helicity and momentum $p$.\footnote{ In Quantum Field Theory, states of particles are characterized by momenta. Because in neutrino production processes it is impossible
to distinguish production of neutrinos with different masses, we assume that mixed neutrino states are also characterized by definite momenta. Let us stress that in the flavor neutrino state formalism we are considering here, this assumption  is the only possibility to get  the standard oscillation phases which are in agreement with the data of all neutrino oscillation experiments. In this sense our assumption is confirmed by experiments.
 We will also notice that the theory of the evolution of neutrinos in matter and the MSW effect \cite{Wolf,MS} are based on the assumption that a flavor neutrino state is a state with definite momentum.}

 Let us stress that
\begin{itemize}
  \item Flavor neutrino states do not depend on the production process.
  \item Flavor states are characterized by the momentum (if there are no special conditions of neutrino production).
\item Flavor states are orthogonal and normalized
\begin{equation}\label{decay13}
\langle \nu_{l'}|\nu_{l}\rangle=\delta_{l'l}.
\end{equation}
\end{itemize}
The evolution of states in QFT is given by the Schr\"odinger equation
\begin{equation}\label{Schrod}
i\frac{\partial~ |\Psi(t)\rangle}{\partial t}=H~|\Psi(t)\rangle,
\end{equation}
where $H$ is the total Hamiltonian and time $t$ is a parameter both of which characterize the evolution of the system.

If at $t=0$ in a CC weak process $\nu_{l}$ is  produced, we have
 for the  state of the neutrino   at the time $t$
\begin{equation}\label{Schrod1}
|\nu_{l}\rangle_{t}=e^{-iHt}|\nu_{l}\rangle=
\sum_{i}|\nu_{i}\rangle e^{-iE_{i}t}~U^{*}_{li},
\end{equation}
where
\begin{equation}\label{Schrod2}
H|\nu_{i}\rangle =E_{i}|\nu_{i}\rangle,\quad E_{i}\simeq E+ \frac{m^{2}_{i}}{2E}.
\end{equation}

Neutrinos are detected via the observation of weak CC and NC processes. Let us consider the production of a lepton $l'$ in the CC process
\begin{equation}\label{detect}
\nu_{i}+N\to l' +X.
\end{equation}
Taking into account that effects of neutrino masses can not be resolved in neutrino processes we have
\begin{equation}\label{detect1}
\langle l' X|S|\nu_{i}N \rangle\simeq
\langle l' X|S|\nu_{l'}N\rangle_{SM}~U_{l'i},
\end{equation}
where $\langle l' ~X|S|\nu_{l'}~N\rangle_{SM}$ is the SM matrix element of the process
\begin{equation}\label{detect2}
\nu_{l'}+N\to l' +X.
\end{equation}
From (\ref{matelem5}), (\ref{Schrod1}) and (\ref{detect1}) follows that the chain  of processes $a\to b +l^{+}+\nu_{l}$,~~ $\nu_{l}\to \nu_{l'}$,
~~$\nu_{l'}+N\to l'+X$ corresponds to the following {\em factorized product} of amplitudes
\begin{equation}\label{detect3}
\langle l' ~X|S|\nu_{l'}~N\rangle_{SM}~
\left(\sum_{i}U_{l'i}~e^{-iE_{i}t}~U^{*}_{li}\right)~\langle b ~l^{+}\nu_{l}|S|a\rangle_{SM}.
\end{equation}
Only the amplitude of the transition $\nu_{l}\to \nu_{l'}$
\begin{equation}\label{detect4}
\mathcal{A}(\nu_{l}\to \nu_{l'})=\sum_{i}U_{l'i}~e^{-iE_{i}t}~U^{*}_{li}
\end{equation}
depends on the properties of massive neutrinos (mass-squared differences and mixing angles). The matrix elements of the neutrino production and detection are given by the Standard Model expressions in which effects of neutrino masses can safely be neglected.
Let us stress that the property of the factorization (\ref{detect3}) is based on the smallness of the neutrino masses and on the Heisenberg uncertainty relation.

\section{Mandelstam-Tamm uncertainty relation\\
and neutrino oscillations}

All uncertainty relations in Quantum Theory are based on the inequality
\begin{equation}\label{uncertain}
\Delta A~\Delta B
\geq \frac{1}{2}|\langle a|[A,B]|a\rangle|
\end{equation}
which follows from the Cauchy inequality. In (\ref{uncertain}) $A$ and $B$ are hermitian operators, $|a\rangle$ is any state, $\Delta A=\sqrt{\langle a|(A-\overline{A})^{2}|a\rangle}$ is the  standard deviation and
$\overline{A}=\langle a|A|a\rangle$ is the average value of the operator $A$. For example, for operators of momentum $p$ and coordinate $q$ which satisfy the commutation relation
$[p,q]=\frac{1}{i}$ we have
the Heisenberg uncertainty relation $\Delta p~\Delta q\geq \frac{1}{2}$.

The Mandelstam-Tamm time-energy uncertainty relation \cite{MandTamm} is based on
the inequality (\ref{uncertain}) and the equation
\begin{equation}\label{uncertain1}
 i\frac{\partial O(t)}{\partial t}=[O(t),H]
\end{equation}
for any operator  $O(t)$ in the Heisenberg representation ($H$ is the total Hamiltonian).

From (\ref{uncertain}) and (\ref{uncertain1}) we have
\begin{equation}\label{uncertain2}
\Delta E~\Delta O(t)\geq\frac{1}{2}|\frac{d }{d t}\overline{ O}(t)|
\end{equation}
This inequality gives nontrivial constraints only in the
case of {\em non-stationary states}.

Taking into account that $\Delta E$ does not depend on $t$ we find
\begin{equation}\label{uncertain3}
\Delta E ~\Delta t \geq\frac{1}{2}\frac{ |\overline{ O}(\Delta t)-\overline{ O}(0)|}
{\Delta O(\bar t)}
\end{equation}
For the time interval $\Delta t$ during which the state of the system
is significantly changed
($\overline{ O}( t)$ is changed
by the value which is characterized by the standard deviation) the right-hand part of
(\ref{uncertain3}) is of the order of one. We obtain the Mandelstam-Tamm time-energy  uncertainty relation
\begin{equation}\label{uncertain4}
\Delta E ~\Delta t \gtrsim 1.
\end{equation}

From (\ref{detect4}), for the normalized probability of the transition $\nu_{l}\to \nu_{l'}$ we obtain the expression
\begin{equation}\label{Probabil}
P(\nu_{l}\to \nu_{l'})=
|\sum_{i\neq j}U_{l'i}~(e^{-i(E_{i}-E_{j})t}-1)~U^{*}_{li}+\delta_{l'l}|^{2},
\end{equation}
which obviously gives the standard transition probability (\ref{probabil}).

From (\ref{Probabil}) follows that  neutrino oscillations can be observed if the condition
\begin{equation}\label{Probabil1}
|E_{i}-E_{j}|~t \gtrsim 1
\end{equation}
is satisfied.\footnote{This is a necessary condition for the observation of  oscillations. It is also necessary that mixing angles would be relatively large.} It is obvious that this inequality is the Mandelstam-Tamm time-energy uncertainty relation. According to this relation a change of the flavor neutrino state in time  requires energy uncertainty (i.e., a non-stationary state). The time interval required for a significant change of the flavor neutrino state  is given by $t \simeq \frac{1}{|E_{i}-E_{j}|}=\frac{2E}{|\Delta m_{ji}^{2}|}$.\footnote{Let us notice that the inequality (\ref{Probabil1}) can be interpreted in another way:
In order to reveal a small energy difference  $|E_{i}-E_{j}|\simeq \frac{|\Delta m_{ji}^{2}|}{2E}$ we need a large time interval $t \gtrsim \frac{1}{|E_{i}-E_{j}|}$. This corresponds to another interpretation of the time-energy uncertainty relation (see \cite{Fock}).}

\section{On plane wave and wave packet approaches to neutrino oscillations}

We will now briefly discuss other approaches to neutrino oscillations.
In the approach based on the relativistic quantum mechanics, in CC processes together with charged leptons {\bf coherent superpositions of plane waves} are produced and absorbed.
In this case, for the normalized $\nu_{l}\to \nu_{l'}$ transition probability
the following expression can be obtained (see, for example\cite{Levy,Giunti})
\begin{equation}\label{xevolution1}
P(\nu_{l}\to \nu_{l'})=
|\sum_{i}U_{l'i}e^{-ip_{i}\cdot x}U^{*}_{li}|^{2}=|\sum_{i\neq j}U_{l'i}(e^{-i(p_{i}-p_{j})\cdot x}-1)U^{*}_{li}+\delta_{l'l}|^{2}.
\end{equation}
Here $p_{i}=(E_{i},\vec{p}_{i})$ is the 4-momentum of a neutrino with mass $m_{i}$ and
 $x=(t,\vec{x})$.

Let us assume that
$\vec{p}_{i}=p_{i}\vec{k}$, where $\vec{k}$ is the unit vector. For the phase difference
which is gained by a plain wave at the distance $x=(\vec{x}\vec{k})=L$ after the time interval $t$ we have
\begin{equation}\label{xevolution2}
(p_{i}-p_{j})\cdot x = (E_{i}-E_{j})t-(p_{i}-p_{j})L.
\end{equation}
For ultrarelativistic neutrinos we have
\begin{equation}\label{xevolution3}
    t\simeq L.
\end{equation}
Taking into account that $E_{i}\simeq p_{i}+\frac{m_{i}^{2}}{2E}$, from (\ref{xevolution2}) and (\ref{xevolution3}) we come to the standard oscillation phase
\begin{equation}\label{xevolution4}
(p_{i}-p_{j})\cdot x =\frac{\Delta m_{ji}^{2}}{2E}L
\end{equation}
and the standard expression (\ref{probabil1}) for the transition probability.

Let us stress that in the approach based on the QFT Schr\"odinger equation
the small oscillation phase difference is the result of the cancellation of large terms in the expressions for the neutrino energies. The cancellation  takes place because  neutrino states are characterized by definite momentum. In the QM plane wave approach, small oscillation phases are the result of the cancellation of large terms in the time and space parts of the phase difference. The cancellation is due to the relation (\ref{xevolution3}).

A direct generalization of the QM plane wave approach is
{\bf the wave packet approach}
(see \cite{Giunti} and references therein) in which the plane wave transition probability
(\ref{xevolution1})
is changed to
\begin{equation}\label{wavepack}
P(\nu_{l}\to \nu_{l'})=|\sum_{i}U_{l'i}\int e^{i(\vec{p_{i}}'\vec{x}-
E'_{i}t)}
~f(\vec{p_{i}}'-\vec{p_{i}})~d^{3}p'~U^{*}_{li}|^{2},
\end{equation}
where $E'_{i}=\sqrt{(\vec{p_{i}}')^{2}+m^{2}_{i}}$ and the function
$f(\vec{p_{i}}'-\vec{p}_{i})$ has a sharp maximum at the point
$\vec{p_{i}}'=\vec{p}_{i}$.

Expanding $E'_{i}$ at the point $\vec{p_{i}}'=\vec{p}_{i}$ we find
\begin{equation}\label{wavepack3}
\int e^{i(\vec{p_{i}}'\vec{x}-
E'_{i}t)}
~f(\vec{p_{i}}'-\vec{p_{i}})~d^{3}p'= e^{i(\vec{p_{i}}\vec{x}-
E_{i}t)}
~g(\vec{x}-\vec{v_{i}t}),
\end{equation}
where
\begin{equation}\label{wavepack4}
g(\vec{x}-\vec{v}_{i}t)=\int e^{i\vec{q}~(\vec{x}-\vec{v}_{i}t)}~f(\vec{q})~d^{3}q
\end{equation}
and
\begin{equation}\label{wavepack2}
\vec{v}_{i}=\frac{\vec{p}_{i}}{E_{i}},\quad E_{i}=\sqrt{\vec{p_{i}}^{2}+m^{2}_{i}}.
\end{equation}
If we  make the standard assumption that the function $f(\vec{q})$   has the Gaussian form
\begin{equation}\label{wavepack5}
f(\vec{q})=N~e^{-\frac{q^{2}}{4\sigma_{p}^{2}}},
\end{equation}
($\sigma_{p}$ is the width of the wave packet in the momentum space)
we find
\begin{equation}\label{wavepack6}
g(\vec{x}-\vec{v}_{i}t)=N (\frac{\pi}{\sigma_{x}^{2}})^{3/2}~e^
{-\frac{(\vec{x}-\vec{v}_{i}t)^{2}}{4\sigma_{x}^{2}}},
\end{equation}
where $\sigma_{x}=\frac{1}{2\sigma_{p}}$ characterizes the spacial width of the wave packet.

The probability of the transition $\nu_{l}\to \nu_{l'}$ in the wave packet approach is determined as a quantity integrated over  time.
From (\ref{wavepack3}) we find the following expression for the integrated
normalized transition probability
\begin{equation}\label{wavepack11}
\mathcal{P}(\nu_{l}\to \nu_{l'})=\sum_{i,k}U_{l'i}U^{*}_{l'k}
e^{i[(p_{i}-p_{k})-(E_{i}-E_{k})]L}U^{*}_{li}U_{lk}~
e^{-(\frac{L}{L^{ik}_{\mathrm{coh}}})^{2}}~e^{-2\pi^{2}\xi^{2}
(\frac{\sigma_{x}}{L^{ik}_{\mathrm{osc}}})^{2}}.
\end{equation}
Here $L$ is the distance between neutrino source and neutrino detector, $L^{ik}_{\mathrm{osc}}$ is the oscillation length,
$\xi$ is a constant of the order of one and
\begin{equation}\label{wavepack10}
L^{ik}_{\mathrm{coh}}=\frac{4\sqrt{2}\sigma_{x}E^{2}}{|\Delta m^{2}_{ik}|}.
\end{equation}
is the coherence length.\footnote{We have $|v_{i}- v_{k}|L^{ik}_{\mathrm{coh}}\simeq \frac{|\Delta m^{2}_{ik}|}{2E^{2}}L^{ik}_{\mathrm{coh}}\sim 2\sqrt{2}\sigma_{x}$. Thus, the
coherence length is such a distance between neutrino source and detector at which  $\nu_{i}$ and
$\nu_{k}$ are separated by an interval comparable to the size of the wave packet.}
Taking into account that $(p_{i}-p_{k})-(E_{i}-E_{k}) = -\frac{\Delta m^{2}_{ki}}{2E}$
we come to the conclusion that the $\nu_{l}\to \nu_{l'}$ transition probability in the wave packet approach is given by the standard expression (\ref{probabil}) which is multiplied
by the decoherence factor $e^{-(\frac{L}{L^{ik}_{\mathrm{coh}}})^{2}}$ and the factor
$e^{-2\pi^{2}\xi^{2}
(\frac{\sigma_{x}}{L^{ik}_{\mathrm{osc}}})^{2}}$.

Thus, the wave packet approach (after integration over $t$) assures the equality $t=L$ and the standard oscillation phase in the transition probability. For usual neutrino oscillation experiments with $L$ being a few times
$L^{A,S}_{\mathrm{osc}}$, the two additional exponential factors
are practically equal to one.

In many papers (see \cite{Oscillations}), neutrinos propagating  about 100 km (reactor $\bar\nu$'s ) or  about 1000 km (atmospheric and accelerator $\nu$'s ), are considered as {\bf virtual particles} in a Feynman diagram-like picture with the neutrino production process at one vertex and the neutrino absorption process in another vertex.
This approach gives the wave packet picture of neutrino oscillations with a transition probability which (before integration over $t$) depends on $x$ and $t$.

The major difference between different approaches to neutrino oscillations can be summarized as follows:
\begin{enumerate}
  \item The QFT approach with the Schr\"odinger evolution equation is based on the assumption of the existence of "mixed" flavor neutrinos $\nu_{e}, \nu_{\mu}, \nu_{\tau}$ which are described by coherent states $|\nu_{l}\rangle=\sum_{i}U^{*}_{li}|\nu_{i}\rangle$. The important characteristic feature of this approach is the Mandelstam-Tamm time-energy uncertainty relation. Neutrino oscillations can take place only in the case of  non-stationary neutrino states with $\Delta E\Delta t\gtrsim 1$, where $\Delta t $ is the time interval during which the oscillations happen. The QFT approach is based on the same general principles as the approach to $K^{0}\rightleftarrows \bar K^{0}$,~$B^{0}\rightleftarrows \bar B^{0}$, etc. oscillations studied in detail at B-factories and other facilities.
  \item
 Other approaches are based on the assumption that in weak processes, mixed coherent superpositions of plane waves or wave packets describing neutrinos with different masses, are produced and detected. The evolution of mixed neutrino wave functions in space and time is determined by the Dirac equation. There is no notion of flavor neutrino states in these approaches. Neutrino oscillations are possible also in the case of monochromatic neutrinos.
\end{enumerate}
Different approaches to neutrino oscillations lead to the same expression for the neutrino transition probability $\mathcal{P}(\nu_{l}\to \nu_{l'})$ in the standard neutrino oscillation experiments. In order to distinguish 1. and 2.  special neutrino oscillation experiments are necessary. Such experiments could be
M\"ossbauer neutrino experiments which we will discuss in the next sections.

\section{M\"ossbauer $\bar\nu_{e}$: Basic considerations}

The basic concept is to use electron antineutrinos ($\bar\nu_{e}$) which are emitted {\em without recoil} in a bound-state $\beta$-decay and are resonantly captured {\em again without recoil} in the reverse bound-state process. As an example, let us consider the $^{3}$H - $^{3}$He system \cite{Kells} with the transitions
$\smallskip$

$^{3}$H$\to^{3}$He $+ \bar\nu_{e}$ ~~(source)~~~and~~$\bar\nu_{e} + ^{3}$He $\to^{3}$H~~(target).
$\smallskip$

In the source, the electron  ($e^{-}$)  is emitted directly into a bound-state atomic orbit of $^{3}$He. This decay is a two-body process, thus the emitted $\bar\nu_{e}$ has a fixed energy (18.6 keV). In the target the reverse process occurs, a monochromatic $\bar\nu_{e}$ with an energy of 18.6 keV and an $e^{-}$ in an atomic orbit of  $^{3}$He are absorbed to form  $^{3}$H.

To suppress thermal motions of the  $^{3}$H and  $^{3}$He atoms, they have to be imbedded in a solid-state lattice, e.g., in Nb metal \cite{RajuRag}. In addition, for a M\"ossbauer  $\bar\nu_{e}$ experiment it is mandatory that no phonons are excited in the lattice when the  $\bar\nu_{e}$  is emitted or absorbed, because only then a highly monochromatic  $\bar\nu_{e}$ radiation and the large cross section of the M\"ossbauer resonance of typically $10^{-19}$ to  $10^{-17}$ cm$^{2}$ can be achieved. However, it became apparent \cite{PotzelWagnerPRL},\cite{Schiffer},\cite{PotzelAlushta},\cite{PotzelPoland} that there exist several basic difficulties to observe  M\"ossbauer  $\bar\nu_{e}$ with the system $^{3}$H - $^{3}$He in Nb metal. The main problem originates from lattice expansion and contraction processes. They occur when the nuclear transformations (from  $^{3}$H to  $^{3}$He and from  $^{3}$He to  $^{3}$H) take place during which the $\bar\nu_{e}$ is emitted or absorbed and can cause lattice excitations (phonons) which change the $\bar\nu_{e}$ energy and thus destroy the M\"ossbauer resonance. It has been estimated that due to these lattice excitations the probability for phononless emission and consecutive phononless capture of  $\bar\nu_{e}$ is $\sim7\cdot10^{-8}$ which makes a real experiment with the $^{3}$H - $^{3}$He system extremely difficult \cite{PotzelWagnerPRL},\cite{Schiffer},\cite{PotzelAlushta},\cite{PotzelPoland}. Another basic problem is caused by inhomogeneities in an imperfect lattice which {\em directly} influence the energy of the $\bar\nu_{e}$ \cite{PotzelWagnerPRL}.

 A promising alternative is the rare-earth system  $^{163}$Ho - $^{163}$Dy. It offers several advantages: Due to the highly similar chemical behaviour of the rare earths also the lattice deformation energies for $^{163}$Ho and $^{163}$Dy can be expected to be similar, thus leaving the $\bar\nu_{e}$ energy practically unchanged. In addition, the $\bar\nu_{e}$ energy is very low (2.6 keV), i.e., the recoil originating from the emitted (absorbed) $\bar\nu_{e}$ is highly unlikely to generate phonons in the lattice. Altogether, the probability of phononless emission and absorption could be larger than for the  $^{3}$H - $^{3}$He system by $\sim7$ orders of magnitude. Furthermore, due to the similar chemical behaviour, the $^{163}$Ho - $^{163}$Dy system can also be expected to be less sensitive to variations of the binding energies in the lattice. For this reason, variations of the $\bar\nu_{e}$ energy will also be reduced improving the monochromaticity (linewidth) of the  $\bar\nu_{e}$ M\"ossbauer resonance.

On the negative side, the magnetic moments of the 4f electrons of the rare-earth atoms are large and might cause broadening of the M\"ossbauer  $\bar\nu_{e}$ resonance \cite{PotzelAlushta},\cite{PotzelPoland}. Fortunately, conventional M\"ossbauer spectroscopy (with photons) gathered a wealth of information on the behaviour of rare-earth systems in the past. Of particular interest is the 25.65 keV M\"ossbauer resonance in  $^{161}$Dy where an experimental linewidth of $\Gamma_{exp}\approx5\cdot10^{-8}$ eV has been reached \cite{Greenwood},\cite{PotzelAlushta}. We will show in the following section that the $^{163}$Ho - $^{163}$Dy system might be suitable to investigate the question concerning the different approaches to neutrino oscillations.

\section{The $^{163}$Ho - $^{163}$Dy M\"ossbauer system and the evolution of the $\bar\nu_{e}$ state in time}

If the evolution of the $\bar\nu_{e}$ state occurs in time only, M\"ossbauer $\bar\nu_{e}$ oscillations with an oscillation length $L^{A}_\mathrm{osc}$ determined by  $\Delta m^{2}_{A}$ will \textit{not} be observed if the relative energy uncertainty fulfills the relation \cite{Bilenkyetal2}

\begin{equation}\label{energy uncertainty}
\frac{\Delta E}{E}\ll\frac{1}{4}\frac{\Delta m_{A}^{2}c^{4}}{E^{2}}
\end{equation}

where $\Delta m_{A}^{2}\simeq 2.4\cdot 10^{-3}~\mathrm{eV}^{2}$ is the atmospheric mass-squared difference. For the $^{163}$Ho - $^{163}$Dy system, eq. (\ref{energy uncertainty}) requires $\left(\frac{\Delta E}{E}\right)_{Ho-Dy}\ll 9.2\cdot10^{-11}$ or $\Delta E\ll 2.4\cdot10^{-7}$ eV.

For the 25.65 keV $\gamma$-transition in $^{161}$Dy an experimental linewidth of $\Gamma_{exp}\approx5\cdot10^{-8}$ eV has been observed \cite{Greenwood}, which is $\sim5$ times below the limit $\Delta E\lesssim 2.4\cdot10^{-7}$ eV just mentioned. It might be expected that a similar value for $\Gamma_{exp}$ can be reached for the $^{163}$Ho - $^{163}$Dy system. In particular, using the usual M\"ossbauer $\gamma$-transition in $^{161}$Dy, relevant physical properties, e.g., the experimental linewidth in the Ho - Dy system can be investigated and improved if necessary. Thus it looks promising that the question  if M\"ossbauer $\bar\nu_{e}$ oscillate can be answered experimentally. For $\Gamma_{exp}\approx5\cdot10^{-8}$ eV, according to the Mandelstam-Tamm time-energy uncertainty relation a significant change of the $\bar\nu_{e}$ state in time can occur only very slowly leading to a long oscillation path-length $L_{change}$ since the $\bar\nu_{e}$ is ultrarelativistic:

\begin{equation}\label{changeduringtime}
L_{change}\simeq c\cdot\frac{\hbar}{\Gamma_{exp}}\cdot 2\pi.
\end{equation}

For the $^{163}$Ho - $^{163}$Dy system, $L_{change}\approx25$ m for the $\bar\nu_{e}$ state.

In comparison, for an evolution of the $\bar\nu_{e}$ state in space and time, the oscillation length is given by eq. (\ref{osclength}). With $E=2.6$ keV for the $^{163}$Ho - $^{163}$Dy system, and $\Delta m_{A}^{2}\simeq 2.4\cdot 10^{-3}~\mathrm{eV}^{2}$, we obtain $L^{A}_\mathrm{osc}\simeq2.6$ m, about 10 times shorter than $L_{change}$. If the evolution occurs in time only, in such a M\"ossbauer-neutrino experiment with $\Gamma_{exp}\approx5\cdot10^{-8}$ eV, instead of $L^{A}_\mathrm{osc}$ the much longer $L_{change}$ would be observed.

If M\"ossbauer $\bar\nu_{e}$ oscillate, an interesting application would be the search for the conversion to sterile neutrinos  $\bar\nu_{e}\to \bar\nu_{sterile}$ \cite{Kopeikin} involving additional mass eigenstates. Since  $\bar\nu_{sterile}$ does not show the weak interaction of the Standard Model of elementary particle interactions, such a conversion would have to be tested by the disappearance of  $\bar\nu_{e}$. The results of the LSND (Liquid Scintillator Neutrino Detector) experiment \cite{LSND},\cite{LSND1} indicate a mass splitting of $\Delta m^{2}\approx1$ eV$^2$ \cite{RajuRag}. Unfortunately, several experiments performed by the MiniBooNE collaboration to check the LSND results have not been conclusive, although the MiniBooNE results are compatible with the LSND observation \cite{MiniBooNE}. For M\"ossbauer $\bar\nu_{e}$ of the  $^{163}$Ho - $^{163}$Dy system ($E$=2.6 keV) the oscillation length $L^{A}_\mathrm{osc}$ would be only $\sim1$ cm if  $\Delta m^{2}\approx1$ eV$^2$.

\section{Conclusions}
After the golden years of the discovery of neutrino oscillations in atmospheric, solar and reactor neutrino experiments we now enter into the era of detailed studies of this phenomenon. Measurements of the small mixing angle $\theta_{13}$, of the $CP$ phase $\delta$, and the establishment of the character of the neutrino-mass spectrum will require high-precision neutrino-oscillation experiments which are already ongoing now or are under preparation or in the R\&D stage.

Is there a consensus in the treatment and understanding of the neutrino oscillation phenomenon? Many recent papers on the theory of neutrino oscillations (see, for example, \cite{Oscillations}) certify that such a consensus still does not exist.

Is  the notion of flavor neutrinos $\nu_{e}$, $\nu_{\mu}$ and $\nu_{\tau}$ in the case of neutrino mixing a convenient terminology coming from "the times of massless neutrinos" or are they real physical states? From the momentum-coordinate Heisenberg uncertainty relation follows that due to the small values of the neutrino mass-squared differences in weak processes "mixed" flavor neutrinos $\nu_{l}$ (similar to the "mixed" $K^{0}$, $\bar K^{0}$;
$B^{0}$, $\bar B^{0}$; etc), which are described by coherent superpositions of states of neutrinos with definite mass, are produced and detected. We showed that in this approach for neutrino oscillations to be observed the Mandelstam-Tamm time-energy uncertainty relation must be satisfied. This means that neutrino oscillations can take place only in the case of non-stationary neutrino states.

We compared different approaches to neutrino oscillations. In approaches in which flavor neutrinos are described by coherent superpositions of plane waves or wave packets and in the approach in which neutrinos are considered as virtual particles in a Feynman diagram
with the neutrino production process at one vertex and the neutrino absorption process in another vertex
neutrino oscillations are possible also in the case of monochromatic neutrinos (M\"ossbauer neutrinos).

Usual neutrino oscillation experiments do not allow to distinguish these different approaches. The realization of an idea concerning the M\"ossbauer resonance neutrino experiment with practically monoenergetic $\bar \nu_{e}$ could be the way of probing the
real nature of mixed flavor states, different conjectures on the evolution of such states and the universal applicability of the time-energy uncertainty relation.
 Such an experiment was discussed for the
$^{3}\mathrm{H}-^{3}\mathrm{He}$ source-detector pair \cite{RajuRag}. Recently, however, it was shown that the performance of a M\"ossbauer neutrino experiment in the case of the $^{3}\mathrm{H}-^{3}\mathrm{He}$ system is most probably not possible in practice \cite{PotzelWagnerPRL},\cite{PotzelAlushta},\cite{PotzelPoland}. We present here arguments in favor of a M\"ossbauer neutrino experiment with the
$^{163}\mathrm{Ho}-^{163}\mathrm{Dy}$ source-detector system \cite{PotzelAlushta}. The possibility to perform an experiment in such a system looks promising but is still very challenging and requires further investigations.

\textbf{Acknowledgments}\\
This work was supported by funds of the Deutsche For\-schungsgemeinschaft DFG (Transregio 27: Neutrinos and Beyond), the Munich Cluster of Excellence (Origin and Structure of the Universe), and the  Maier-Leibnitz-Laboratorium (Garching).

\end{document}